\documentclass[12pt]{article}
\usepackage{amsmath,euscript,amssymb}
\usepackage[cp866]{inputenc}
\usepackage[T2A]{fontenc}
\usepackage[english]{babel}
\usepackage{graphicx}
\usepackage{bm}
\newcommand{\be}{\begin{equation}}
\newcommand{\ee}{\end{equation}}

\newcommand{\bea}{\begin{eqnarray}}
\newcommand{\eea}{\end{eqnarray}}

\begin{document}

\title{On the investigations of galaxy redshift periodicity}
\author{K. Bajan,
 P. Flin, W. Godlowski \footnote{1) Astronomical Observatory of
the Jagiellonian University, ul. Orla 171, 30-244 Krakow, Poland},
V.P. Pervushin\footnote{2)Bogoliubov Laboratory of Theoretical
Physics, Joint Institute for Nuclear Research, Dubna, Moscow
Region 141980,
Russia}\\
Pedagogical University, Institute of Physics,\\
ul.Swietokrzyska~15, 25-406 Kielce, Poland}
 \maketitle

%УДК 539.12.01  ~~~~~        PACS~98.65.-r~~~~~ 01.65 +G

\begin{abstract}

In this article we present a historical review of study of the
redshift periodicity of galaxies, starting from the first works
performed in the seventies of the twentieth century until the
present day. We discuss the observational data and methods used,
showing in which cases the discretization of redshifts was
observed. We conclude that galaxy redshift periodisation is an
effect which can really exist. We also discussed the redshift
discretization in two different structures: the Local Group of
galaxies and the Hercules Supercluster. Contrary to the previous
studies we consider all galaxies which can be regarded as a
structure member disregarding the accuracy of velocity
measurements. We applied the power spectrum analysis using the
Hann function for weighting, together with the jackknife error
estimator. In both the structures we found weak effects of
redshift periodisation.

\end{abstract}

\section{Introduction}

In the large scale Universe, the search for regulations is
connected with testing  radial velocities of galaxies and quasars.
We can describe redshift as:

$$z=\frac{\lambda -
\lambda_{1}}{\lambda_{1}}=\frac{R(t_{0})}{R(t_{1})}-1\simeq
\frac{v_{r}}{c}$$ where $\lambda$ is the observed wavelength,
$\lambda_{1}$ is the emitted wavelength, and $R(t)$ is the scale
factor. Redshift depends on:
\begin{enumerate}
\item General expansion of the Universe (Hubble flow)
\item Local peculiarities due to matter distribution
\item Small scale motion of matter inside a galaxy
\end{enumerate}
It is commonly accepted that a radial velocity of an object does
not depend on its position on the celestial sphere, magnitude, and
other properties of  objects.

These velocities can have an arbitrary value or they can be
grouped around some particular values. Any distribution of galaxy
velocities can be described using a continuous or discrete
function. In the first case, it means that redshift can have
arbitrary value. It is possible to find maxima and minima, which
means that some redshift values are more probable than others.
There are maxima in the distribution of object redshifts, which
are separated by a constant value. Such a distribution is called,
not very correctly, redshift discretization. The second case is
the exact discrete distribution. Radial velocities of galaxies can
have only a discrete value. This is a strict quantization of
radial velocities. If sometimes an object with redshift, which is
not a strict multiplication of a periodization value is observed,
this is due to observational errors. Both the above-mentioned
possibilities are called periodisation or discretization. The
discretization of redshift for astronomical objects can be
discussed independently for three cases, namely galaxies, quasars
and large - scale periodicity (about $120~Mpc$). In our previous
paper \cite{bajan1} the quest for quasar redshift periodicity is
described, while the latter studies have been discussed
\cite{bajan2}, too.

In the present paper, we present the investigations of radial
velocities of galaxies.

The subject of redshift periodisation is not very popular,
sometimes even regarded as scientifically suspicious. However, we
share the opinion expressed by Hawkins et al. \cite{hawkins} that
all these effects should be carefully checked. They wrote: "The
criticism usually leveled at this kind of study is that the
samples of redshifts tended to be rather small and were selected
in a heterogeneous manner, which makes it hard to assess their
significance. More cynical critics also point out that the results
tend to come from a relatively small group of astronomers who have
a strong prejudice in favour of detecting such unconventional
phenomena. This small group of astronomers, not unreasonably,
respond by pointing out that adherents to the conventional
cosmological paradigm have at least as strong a prejudice towards
denying such results.

We have attempted to carry out this analysis without prejudice.
Indeed, we would  be happy with either outcome: if the periodicity
were detected, then there would be some fascinating new
astrophysics for us to explore; if it were not detected, then we
would have the reassurance that our existing work on redshift
surveys, ect, has not been based on false premises."

We think that for better understanding of the subject it is
necessary to give a review of obtained results, together with the
manner of their receiving. Therefore, in this paper we present
this story.

\section{The relation between galaxy redshift and magnitude}

First studies on relation between redshift, morphological type and
magnitude of the galaxy core were carried out by Tifft in 1972
\cite{tifft1, tifft2} in the Coma cluster of galaxies. They showed
that galaxies lie in narrow bands on the magnitude redshift
diagram, which slope down in the direction of smaller magnitudes
at higher redshifts. A year later Tifft \cite{tifft3} analyzed a
sample of about 100 galaxies situated in the center of the Coma
cluster. He divided this sample into two smaller ones: first
contained elliptical galaxies only; while the second,
nonelliptical ones. The redshift - magnitude diagram  was
reanalyzed showing a strong band structure. On this diagram there
were 70 points from which 57 were located within the range of
first three bands in the ratio of: $21:18:18$. The observed effect
was compared with another, more distant group of galaxies situated
in the Coma cluster. In the group, almost identical properties but
a bit shifted in redshift were observed. A similar effect occurs
in the $(m,\log z)$ diagram for field galaxies. The statistical
significance of the band structure was checked using the
$\chi^{2}$ test. In 1972, no physical mechanism responsible for
relationship between redshift and magnitude or redshift and
morphology was known, so the attitude to the problem was
completely empirical.

Continuing the Coma cluster investigation, Tifft added to his
sample dimmed objects lying on a bigger area on the sky. In this
way, he obtained a new sample of galaxies containing 108 objects.
The band structure was observed, there were 89 objects situated on
the first three bands. This diagram suggests that galaxies are
situated in subgroups along the lines. In this paper,  the power
spectrum analysis was applied for the first time.

The existence of the band structure has been shown in the
following clusters of galaxies: Coma (contained 108 objects),
A2199 cluster of galaxies (33 objects) and Perseus (90 objects)
\cite{tifft1, tifft3}. However, the idea about convergent bands
has been considered only for the Coma cluster of galaxies
\cite{tifft4}.

\begin{figure}[htb]
\includegraphics[width=10cm]{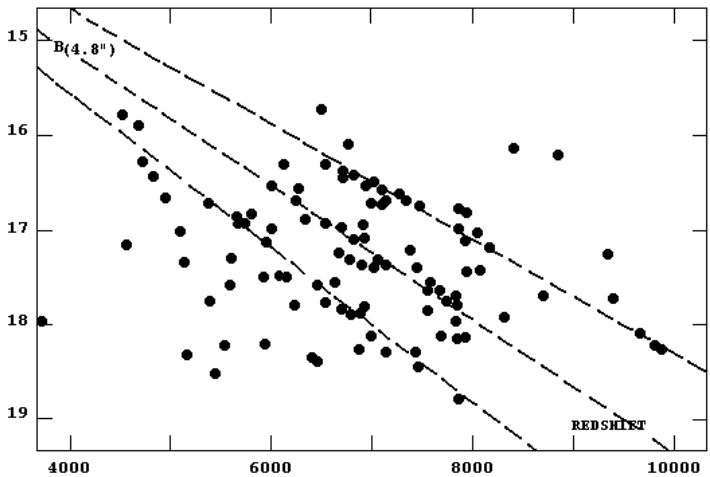}
\caption{Nuclear blue magnitudes $B(4.^{''}8)$ versus redshift in
$km\cdot s^{-1}$. Taken from Nanni et al. \cite{nanni}. With kind
permission of authors, $\copyright$ Astron. Astrophys. (1981)}
\label{fig1}
\end{figure}

Nanni et al. \cite{nanni} reanalyzed the centre of the Coma
cluster of galaxies. Redshifts were taken from Tifft's article
\cite{tifft3}, magnitude from their own observations. They
confirmed the existence of the effect, if seen from the point of
convergence of the bands. But there was no effect in a direction
transversal to the bands. They stated the statistical significance
of these bands. They claimed that this effect could be connected
with the systematical errors in the redshift determinations or
with some dynamical reasons.

The above-mentioned  investigations gave evidences that galaxy
redshift is not an independent observable, it depends on some
previously discarded factors.

\section{First observational evidences of redshift discretization}

A simple model of rotating galaxy is disturbed through noncircular
motions and other phenomena. Tifft \cite{tifft5, tifft6} discussed
the influence of these effects on observed radial velocities of
galaxies. The precision was the main problem in the early
observations of radial velocities of galaxies, because it was too
small for testing the effects postulated by Tifft. He
\cite{tifft5} analyzed the spiral galaxies NGC~2903,
M51~=~NGC~5194 and M31, noticing the existence of the areas with a
difference of about $75~km\cdot s^{-1}$. Moreover, the redshift
differences between the neighbouring galaxies seemed to be the
multiplication of $70~-~75~km\cdot s^{-1}$. Therefore, the
difference in redshift among pair galaxies started to be the main
target of studies.

There are many articles which studied the redshift difference in
the double galaxies. The authors considered newer and better
catalogues of galaxies. Fundamental progress was connected with
both accuracy of velocity measurement and an increasing number of
considered pairs.

In 1976, Turner \cite{turner} published a new catalogue with data
obtained in an optical way. On its base, Tifft \cite{tifft6} carried
a new analysis of all pairs with accuracy of redshifts
determinations $<~100~km\cdot s^{-1}$. The result is statistically
significant: he obtained periodicity in redshift distribution with a
period $72~km\cdot s^{-1}$, $144~km\cdot s^{-1}$ and $216~km\cdot
s^{-1}$. In 1979, the new data appeared: Peterson's catalogue
\cite{peterson}. Almost all of the data received in radio way had
uncertainties of measurement about $20~km\cdot s^{-1}$. Tifft
\cite{tifft7} considered pairs of galaxies with redshift difference
$<~250~km\cdot s^{-1}$ and uncertainty $<~50~km\cdot s^{-1}$. He
obtained strong periodicity with a period $72 \cdot n$ (where
$n=1,2,3Е$) for the whole sample.

In 1982, Tifft \cite{tifft8} working with double galaxies introduced
a few new criteria which led to a sample containing 40 galaxies out
of 48 objects. He concluded that in the region around
$72\pm18~km\cdot s^{-1}$ there were three times pairs more than
outside it. The statistical significance of this result is $99.8\%$
using the $\chi^{2}$ test. There were also populations of small
peaks around $36~km\cdot s^{-1}$ and the ''zero'' peak was shifted
to $12~km\cdot s^{-1}$. Unfortunately, grouping was tested in
$36~km\cdot s^{-1}$ width ranges which was half the searched
periodicity.

In 1982 Tifft \cite{tifft12} finished his catalogue of pairs of
galaxies. He  discovered periodicity for 200 pairs at the
significance level of $99\%$ or higher.

\section{The establishment of the redshift distribution}

In the eighties of the twentieth century the dynamical development
of observational technique was noted. The number of galaxies with
known radial velocities increased as well as the precision of
determining this value. From the point of view of data analysis it
was not interesting time, because the PSA method was established
as the best method used for studying redshift periodicity.
Objections that were put against the method by Newman, Haynes and
Terzian \cite{newman}  found an answer in Cocke and Tifft's
\cite{cocke} work. The second method used for testing periodicity
was the Bernoulli test. So, it started to be clear that possible
discovering of periodicity  is not a result of  method applied;
therefore, the majority of studies were concentrated on data
analysis.

The heliocentric radial velocity of galaxy should be corrected
relative to the Sun's motion around the Galaxy center and/or
around the center of Local Group. In our paper \cite{godlowski},
we applied 11 different corrections taken from different authors.
For example:
\begin{enumerate}
\item the galactocentric reduction: $v=232~km\cdot s^{-1}$, $l=88^{o}$, $b=2^{o}$ \cite{guthrie1}
\item the galactocentric reduction: $v=213~km\cdot s^{-1}\pm10~km\cdot s^{-1}$,
$l=93^{o}\pm3^{o}$, $b=2^{o}\pm5^{o}$ \cite{guthrie3}
\item the pure heliocentric reduction ($v=0~km\cdot s^{-1}$, $l=0^{o}$, $b=0^{o}$)
\item the velocity obtained from the Sun's motion relative to the LG center:
$v=306~km\cdot s^{-1}\pm18~km\cdot s^{-1}$,$l=99^{o}\pm5^{o}$,
$b=-3.5^{o}\pm4^{o}$ \cite{courteau}
\item the cosmocentric reduction: $v=369~km\cdot s^{-1}$, $l=264.7^{o}$, $b=48.2^{o}$
\end{enumerate}
Moreover, it was discovered that the effect of redshift
discretization appeared not only in galactocentric frame of
reference but also in CMB (correction number 5).

The periodicity was  observed only in tha case of galactocentric
radial velocities or using CMB reference frame, not in the case of
heliocentric radial velocity.

In this period researchers had access to a few data catalogues:
Petersons catalogue \cite{peterson}, Haynes data's \cite{haynes}
and Helou, Salpeter and Terzian \cite{helou} data's. They began
using corrections to the Sun's motion \cite{tifft9, croasdale}.
The Monte Carlo simulations and Kolmogorov - Smirnov \cite{sharp}
test  joined to PSA method. The periodicity was searched in the
whole samples \cite{cocke1} or in the subsamples \cite{croasdale}.
The periodicy around values: $24,15~km\cdot s^{-1}$ \cite{tifft9,
tifft10}, $36,3~km\cdot s^{-1}$ \cite{cocke1, croasdale, tifft10},
$72~km\cdot s^{-1}$ \cite{cocke1}, $144~km\cdot s^{-1}$
\cite{cocke1} and $90~km\cdot s^{-1}$ \cite{cocke1} were stated.

In the nineties of the last century the search for periodicity in
galaxies belonging to structures was started . Guthrie and Napier
\cite{guthrie2, guthrie4} considered 2 samples of galaxies: the
first contained 112 spiral galaxies with redshift $<~3000~km\cdot
s^{-1}$ and the second contained 77 dwarf galaxies. They took into
account galaxies lying near the center of the Virgo cluster. The
 existence of redshift periodicity for dwarf
galaxies was not confirmed. However, for the whole sample of
spiral galaxies they found possible periodicity around
$71,1~km\cdot s^{-1}$. Tifft hypothesis was confirmed at 0.99
significance level under the  assumption that Local Group velocity
towards the Virgo cluster is $100~-~400~km\cdot s^{-1}$. Then,
sample of spiral galaxies was  divided due to its location in high
or low density regions. They found strong periodicity near
$71~km\cdot s^{-1}$ in a subsample of galaxies lying in regions of
low density. There were no quantization in the spirals in high
density regions.

Next sample contained 89 spiral galaxies lying on the Virgo cluster
periphery \cite{guthrie1, guthrie4}. The strong redshift periodicity
around $37.2~km\cdot s^{-1}$ was found. But this periodicity
appeared only if galactocentric redshifts were considered.

Then Guthrie and Napier \cite{napier1} took into account galaxies
lying at the edge of the Local Supercluster. They used the
database compiled by Bottinelli et al. \cite{bottinelli}. After
eliminating these belonging to the Virgo cluster and nonspirals
they obtained a sample contained 247 objects. They found that
redshifts of spiral galaxies were strongly periodic around
$37.5~km\cdot s^{-1}$. It should be pointed out that in all these
investigations from database containing thousands of galaxies only
a small number of them, namely those with very accurate
measurements were taken into account.

A few hypothesis about new physics were considered by them
\cite{napier1}. The redshift periodicity can be explained by the
regularity of the LSC structure or applying the model of
oscillation of physical parameters.

If redshifts were taken as radial velocity, one can suppose that
galaxies are situated in the regular structure and galaxies have
small or negligible peculiar motions and that this structure takes
part in the global expansion of the Universe. This model predicts
correctly the quantization ranges. If the quantization range Q is
given as $Q=H_{0}\cdot d$, where $H_{0}$ is the local Hubble
constant, d is provided scale of the cell, then $Q=37.5~km\cdot
s^{-1}$. In the large scale structure of the Universe the
periodisation was detected by pencil beam observation. This
periodisation can be due to walls and voids between them. It is
possible that within walls some aggregation of galaxies exists. So
the small scale periodisation is due to substructures. Hence the
size of cells can be estimated, comparing this value with light
velocity c. We obtain the value $3/8~Mpc$, i.e. almost $400~kpc$,
which corresponds to the size of compact galaxy groups. It is not
clear if this hypothesis can explain the observed streaming
motions, and due to accidental projection it is unable to explain
the periodicity around the value of $72~km\cdot s^{-1}$ observed
by Tifft for double galaxies.

Let us list other possibilities:
\begin{enumerate}
\item Periodic oscillation of physical constants. The
gravitational constant \cite{morikawa} and the Hubble parameter
\cite{morikawa1} were considered, but the required amplitude of
changes was greater than observed limits. Also the changes of fine
structure constant \cite{hill} and the variability of the electron
mass \cite{hill} were considered. But there is a problem in this
model because periodicity can be smeared out by a peculiar motion of
galaxies. \item The Holmlid \cite{holmlid} explanation is quite
natural. As  the periodicity is observed only after applying the
proper corrections to the Sun's motion,  redshifts can be quantized
before they arrive to the Earth. According to Holmlid, the
quantization process have to occur in the intergalactic space
relatively close to observed galaxy, or in the space between our
Galaxy and observed galaxy. The galactic space is not empty but is
filled by the 'dark matter'. This Rydberg matter is not $\Lambda
CDM$ considered in cosmology. This mass is named the Rydberg matter
by Holmlid. In the laboratory after inducing the Rydberg matter
through the laser radiation the blushift was observed. This shift
became redshift due to Stokes dispersion in the cold Rydberg matter
filling interstellar space. So the galaxy redshift is a result of
interaction between the radiation and the cold Rydberg matter.
Holmlid claimed that observed values of redshift periodisation
($36~km\cdot s^{-1}$, $72~km\cdot s^{-1}$, $144~km\cdot s^{-1}$) are
the natural consequence of the structure of clouds composed from
Rydberg matter. \item In 1996, Tifft \cite{tifft11} considered a few
galaxy samples taken from the Virgo cluster, the Perseus and Cancer
Supercluster regions. He examined these samples for periodicities as
viewed from the cosmic background rest frame. He found strong
periodicities around $72~km \cdot s^{-1}$ and $36~km \cdot s^{-1}$.
He thought that this is global feature. As previously only objects
with very accurate measurements were considered, which drastically
diminished the number of galaxies in the analyzed sample. \item
Lehto \cite{lehto} developed a theoretical model which could predict
periods of redshifts. Lehto described basic properties of matter
using 3-dimensional quantized time. The time unit is Planck time and
is  named (by him) "chronon".
\end{enumerate}
The fourth possibility is important because Tifft stressed that
Lehto consideration gave him theoretical explanation of periodicity.

We investigated the periodicity in two structures: Local Group of
Galaxies and Hercules Supercluster. We discuss the distribution of
radial velocities of galaxies belonging to the Local Group
\cite{godlowski}. The main problem is: which galaxies belong to
the Local Group of Galaxies? We discussed 55 objects taken from
the Irwin's list \cite{irwin} together with 7 galaxies from
Maffei's group. That was the first sample. The second sample
containing 32 objects taken from the van den Bergh's list
\cite{bergh}. On Irwin's and van den Bergh's lists there were 9
and 7 objects without determined redshifts, respectively. So we
took into consideration samples containing 28 and 46 galaxies. The
third sample containing 39 objects called "pure" Irwin's data (the
data without galaxies from Maffei's group and without galaxies
with unknown redshifts). The newest data allow us to include
redshifts for 6 from 9 galaxies with unknown redshifts. So for
further consideration we took 5 samples containing 46, 28, 39, 45
and 34 galaxies.

We applied the power spectrum analysis using the Hann function as
a weighting together with the jackknife error estimation. We
perform the detailed analysis of this approach. The distribution
of galaxy redshift seems to be nonrandom. An excess of galaxies
with radial velocities of $24~km\cdot s^{-1}$ and $36~km\cdot
s^{-1}$ is detected, but the effect is statistically weak. Only a
peak for radial velocities $24~km\cdot s^{-1}$ seems to be
confirmed at the confidence level of $95\%$.

\begin{figure}[htb]
\includegraphics[width=12cm,height=6cm]{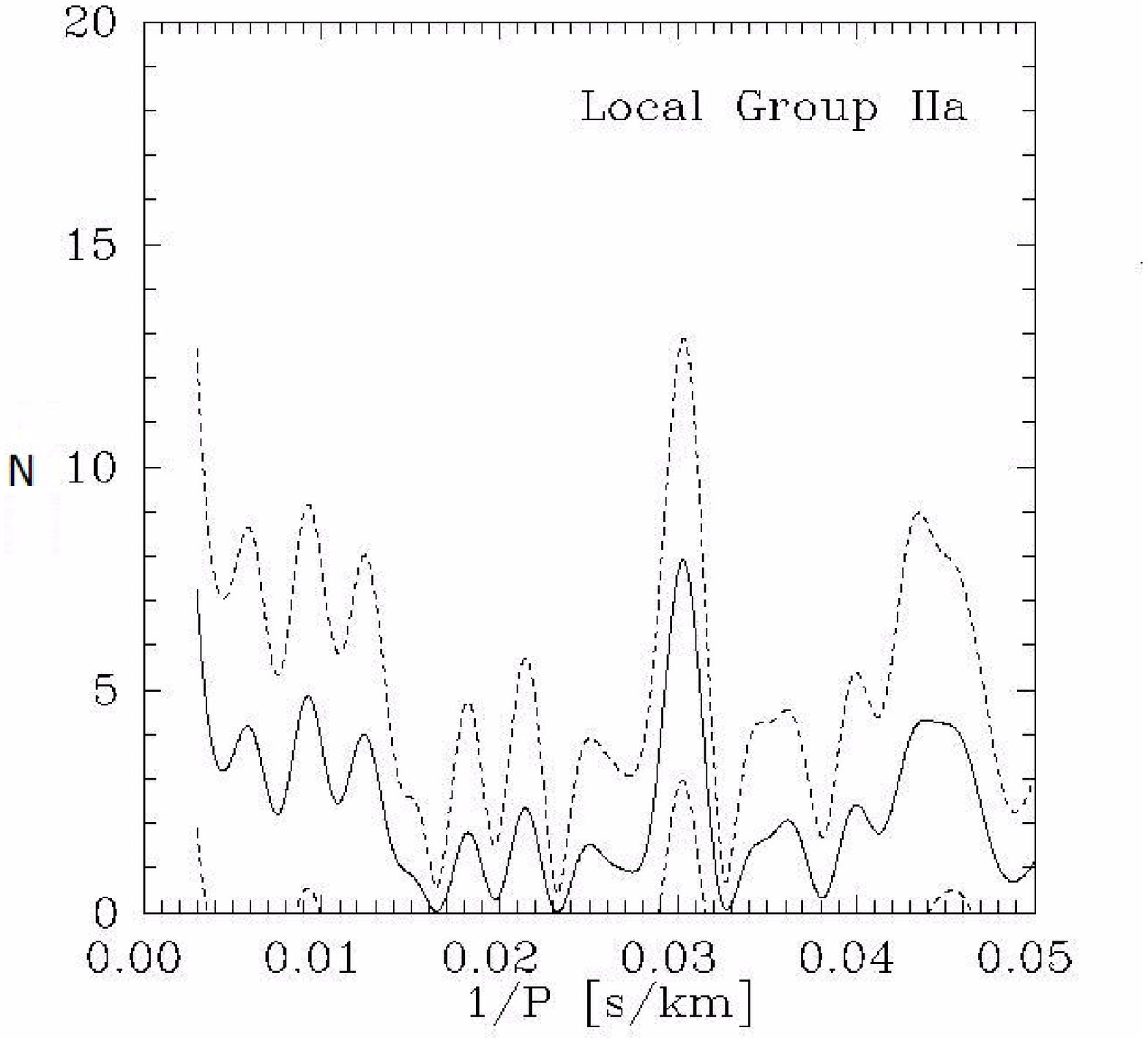}
\caption{The result of PSA analysis for Local Group of galaxies
(sample: 28 galaxies from the van den Bergh list, correction of the
radial velocity: 1) \cite{godlowski}} \label{fig2}
\end{figure}

We also discussed the distribution of radial velocities of
galaxies belonging to the Hercules Supercluster. Our sample
contained 2522 galaxies with radial velocities in the range
$(7500,~15000)~km\cdot s^{-1}$, and it was complete in $80\%$. As
in the above case, we used PSA to analysis that sample. We used 5
velocity corrections enumerated above.

\begin{figure}[htb]
\includegraphics[width=11cm]{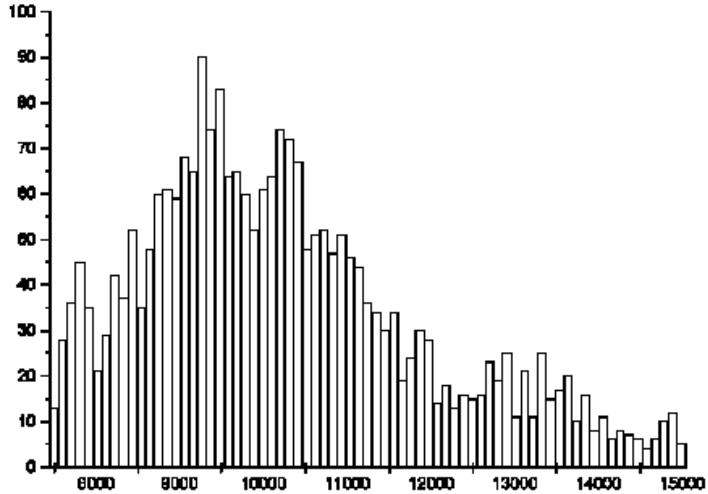}
\caption{The distribution of radial velocities in the Hercules
Supercluster} \label{fig3}
\end{figure}

For galactocentric reduction at the $2\sigma$ confidence level peaks
around $73~km\cdot s^{-1}$ and $24~km\cdot s^{-1}$ are observed.
Although it seemed that the maxima showing in Fig. \ref{fig3} are
clear, the probability that they are coming from nonrandom
distribution is $95\%$, i.e. at the $2\sigma$ significance level.

\begin{figure}[htb]
\includegraphics[width=11cm]{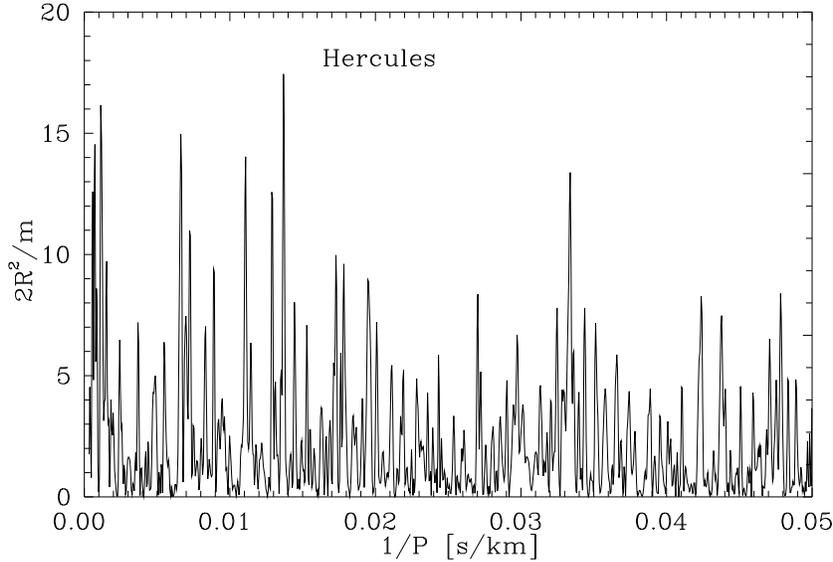}
\caption{The result of the PSA analysis for the Hercules
Supercluster members} \label{fig4}
\end{figure}

\section{Conclusions}
In our opinion the existence of redshift periodicity among
galaxies is not well established. The earlier results are based on
a very small fraction of objects extracted from the large
databases. At the early stage of investigations such an approach
was the correct one, errors of individual measurements were great.
Presently, the radial velocities of galaxies are determined in an
industrial manner. The accuracy of radial velocity determination
is good enough for considering all galaxies. Therefore, we chose
this manner of data treatment. As we considered all galaxies, our
samples are greater. Measurements with lower accuracy could smear
out the regularities, but regularities are not introduced
artificially.

The previous result, based on selected samples showed the
existence of the periodicity in the galaxy redshift distribution
at a very high significance level. We found that at the $2\sigma$
significance level some effect was observed. We think that the
solution of this curious phenomenon can be solved in  near future
using large data base, which together with such correct method as
PSA will allow one to estimate the significance of the effect at a
sufficiently convincing level. We think also that after clear and
convincing demonstrating  of the existence of the effect,
theoretical explanations of this phenomenon can be performed.

\section*{}

\end{document}